\documentclass[usenatbib]{mn2e}
\pdfoutput=1
\usepackage{amssymb}
\usepackage{amsfonts}
\usepackage{amsmath}
\usepackage[pdftex]{graphicx}
\usepackage{epsfig}
\usepackage{color}
\usepackage{dblfloatfix}
\usepackage{float}
\usepackage{fixltx2e}
\usepackage{hyperref}

\def\twid{\mathrel{\lower.1ex\hbox{$\sim$}}}
\def\gtwid{\mathrel{\raise.3ex\hbox{$>$\kern-.75em\lower1ex\hbox{$\sim$}}}}
\def\ltwid{\mathrel{\raise.3ex\hbox{$<$\kern-.75em\lower1ex\hbox{$\sim$}}}}
\def\\{\hfil\break}

\def\hmpc{$h^{-1}$Mpc}

\def\hkpc{$h^{-1}$kpc}
\def\hMsun{$h^{-1}M_{\sun}$}

\def\kmSec{km s$^{-1}$}

\newcommand{\degree}{{\rm o}}

\newcommand{\be}{\begin{equation}}
\newcommand{\ee}{\end{equation}}
\newcommand{\bea}{\begin{eqnarray}}
\newcommand{\eea}{\end{eqnarray}}

\begin{document}

\title[The Cosmic Mach Number]{The Cosmic Mach Number: Comparison from Observations, Numerical Simulations and Nonlinear Predictions}
\author[Agarwal \& Feldman]{
Shankar Agarwal$^{\star}$\,\& Hume A. Feldman$^{\dagger}$\\
Department of Physics \& Astronomy, University of Kansas, Lawrence, KS 66045, USA.\\
emails: $^{\star}$sagarwal@ku.edu; $^{\dagger}$feldman@ku.edu}
\date{} 

\maketitle

\begin{abstract}

We calculate the cosmic Mach number $M$ -- the ratio of the bulk flow of the velocity field on scale $R$ to the velocity dispersion within regions of scale $R$. $M$ is effectively a measure of the ratio of large-scale to small-scale power and can be a useful tool to constrain the cosmological parameter space. Using a compilation of existing peculiar velocity surveys, we calculate $M$ and compare it to that estimated from mock catalogues extracted from the LasDamas (a $\Lambda$CDM cosmology) numerical simulations. We find agreement with expectations for the LasDamas cosmology at $\sim\!1.5\sigma$ CL. We also show that our Mach estimates for the mocks are not biased by  selection function effects. To achieve this, we extract dense and nearly-isotropic distributions using Gaussian selection functions with the same width as the characteristic depth of the real surveys, and show that the Mach numbers estimated from the mocks are very similar to the values based on Gaussian profiles of the corresponding widths. We discuss the importance of the survey window functions in estimating their effective depths. We investigate the nonlinear matter power spectrum interpolator {\sc PkANN} as an alternative to numerical simulations, in the study of Mach number.

\end{abstract}




\section{Introduction}
\label{sec:INTRO}

\cite{OstSut90} introduced a dimensionless statistic of the cosmological structure - the cosmic Mach number, as a way to measure the warmth/coldness of the velocity field on some scale $R$. Specifically, the Mach number is defined as a ratio 
\be
M({\bf x_0};R) \equiv \left(\frac{|{\bf u}({\bf x_0};R)|^2}{\sigma^2({\bf x_0};R)}\right)^{1/2}
\ee
where
${\bf u}({\bf x_0};R)$ is the bulk flow (BF) of a region of size $R$ centered at {$\bf x_0$}, and $\sigma({\bf x_0};R)$ is the velocity dispersion of the objects within this region. The ensemble average over ${\bf x_0}$ gives the statistic $M(R)$. Since both $|{\bf u}({\bf x_0};R)|^2$ and $\sigma^2({\bf x_0};R)$ scale equally by the amplitude of the matter density perturbation, the statistic $M$ is independent (at least in linear approximation) of the normalization of the matter power spectrum. 

In linear theory, given the cosmological parameters, $M$ can be readily calculated and compared with its measured value from the peculiar velocity field catalogues. However, comparing theoretical predictions with observations is not straightforward: (i) one has to correct for the small-scale nonlinearities in observations as well as take into account the fact that observations represent only a discreet sample of the continuous velocity field. This can be remedied by smoothing the velocity field on a suitable scale $r_s$ $(\sim5h^{-1} \textrm{Mpc}$, since on larger scales the matter density field is expected to be linear), before estimating the quantities ${\bf u}({\bf x_0};R)$ and $\sigma({\bf x_0};R)$. However, any residual nonlinearity in the observed field can still bias the $M$ estimates; (ii) Non-uniform, noisy and sparse sampling of the peculiar velocity field can lead to aliasing of small-scale power onto larger scales. When making comparisons with theory, one has to carefully take into account the selection function and the noise of the real dataset. (iii) Peculiar velocity surveys have only line-of-sight velocity information.

Over two decades ago, the statistic $M$ has been investigated in a series of papers: \cite{OstSut90} used linear theory and Gaussian selection function to show that standard Cold Dark Matter (sCDM) model is inconsistent (predicts $M$ almost twice the observed value) with observations at $\sim$ 95\% CL; \cite*{SutCenOst92}, using Tophat and Gaussian selection functions, studied the distribution of $M$ using N-body simulations to rule out the sCDM scenario at 99\% CL; \cite*{StrCenOst93} took into account the selection function of real surveys and extracted mocks from numerical simulations over a range of cosmologies including sCDM and tilted CDM (scalar spectral index, $n_s\neq1$) among others, to reject the sCDM model at 94\% CL. More recently \cite*{MaOstZha12} explored the potential of using $M$ in distinguishing cosmological models, including modified gravity and massive neutrino cosmologies.

In this paper, (i) we estimate the cosmic Mach number for various galaxy peculiar velocity datasets; (ii) we investigate how likely it is to get these Mach values in a $\Lambda$CDM universe. To achieve this, we study the statistical distribution of the expected Mach number by extracting mocks of the real catalogues from numerical simulations of a $\Lambda$CDM universe. We show that a $\Lambda$CDM universe with 7-yr $\it{WMAP}$ type cosmology is consistent with the Mach observations at $\sim\!1.5\sigma$ CL; (iii) we further show that our $M$ estimates for the mocks are not biased by their selection functions. Towards this, we extract dense and nearly isotropic distributions with a Gaussian profile $f(r) \propto e^{-r^2/2R^2}$ with $R=10-100$\hmpc. We show that the Mach numbers estimated from the mocks are very similar to the values based on Gaussian profiles (of similar depth $R$ as the mocks); (iv) we use the nonlinear matter power spectrum interpolation scheme {\sc PkANN} \citep{AgaAbdFelLahTho12} to check if we can avoid N-body simulations completely and predict $M(R)$ by only using {\sc PkANN}'s prediction for the nonlinear power spectrum. This is crucial because high-resolution hydrodynamic N-body simulations are computationally expensive and extremely time consuming. Exploring  parameter space with numerical simulations with reasonable computing resources and time might not be possible. A full use of a statistic like $M$ can only be realized with a prescription for the nonlinear matter power spectrum.

In Sec.~\ref{sec:MACH} we discuss the cosmic Mach number statistic. In Sec.~\ref{sec:SIM} we describe the numerical simulations we use to extract mock surveys. In Sec.~\ref{sec:SURVEYS} we describe the galaxy peculiar velocity surveys (Sec.~\ref{sec:REAL}) and the procedure we follow to extract the mock catalogues (Sec.~\ref{sec:MOCKS}). In Sec.~\ref{sec:MLE} we review the maximum likelihood estimate (hereafter MLE) weighting scheme that is commonly used to analyze peculiar velocity surveys. In Sec.~\ref{sec:STATS}, we show our results for the statistical distribution of the Mach number estimated using various mock catalogues. In Sec.~\ref{sec:PkANN}, we test the performance of {\sc PkANN} - a nonlinear matter power spectrum interpolator, in predicting $M(R)$ for the LasDamas cosmology. The Mach number estimates from real surveys is summarized in Sec.~\ref{sec:MACH_REAL}. We discuss our results and conclude in Sec.~\ref{sec:Conclude}.

\section{The Cosmic Mach Number}
\label{sec:MACH}

Given a peculiar velocity field {\bf v(x)}, one can calculate the bulk flow (see \citealt{FelWat94}, for details), which represent the net streaming motion of a region in some direction relative to the background Hubble expansion. The bulk flow ${\bf u}({\bf x_0};R)$ of a region of size $R$ centered at {$\bf x_0$} can be defined as
\begin{eqnarray}
 \label{eq:bulk_x}
 {\bf u}({\bf x_0};R) &=& \int   d{\bf x} \  {\bf v(x)} F(|{\bf x - x_0}|,R),
\end{eqnarray}
where $F(|{\bf x - x_0}|,R)$ is the filter used to average the velocity field {\bf v(x)} on a characteristic scale $R$. Although Tophat and Gaussian filters are the preferred choices, $F(|{\bf x - x_0}|,R)$ can be designed to mimic the selection function of the real datasets. This is useful when dealing with datasets whose selection function depends strongly on the position in the sky. In Fourier space, Eq.~\ref{eq:bulk_x} can be written as
\begin{eqnarray}
 \label{eq:bulk_k}
 {\bf u}({\bf x_0};R) &=& \int   d{\bf k} \  {\bf v(k)} W({\bf k},R) e^{-i{\bf k \cdot x_0}},
\end{eqnarray}
where ${\bf v(k)}$ and $W({\bf k},R)$ are the Fourier transforms of the peculiar velocity field ${\bf v(x)}$ and the filter $F(|{\bf x - x_0}|,R)$, respectively.

In linear theory of structure formation, at low redshifts, the velocities are related to the matter overdensities via
\begin{eqnarray}
 \label{eq:vk}
 {\bf v(k)} &=& i f H_0 \delta({\bf k}) \frac{\bf k}{k^2},
\end{eqnarray}
where $H_0$ is the present-day Hubble parameter in units of km s$^{-1}$ Mpc$^{-1}$; $\delta({\bf k})$ is the Fourier transform of the overdensity field $\delta({\bf x})$; the linear growth rate factor $f$ can be approximated as $f=\Omega_m^{0.55}$ \citep{Lin05}. Thus, the velocity power spectrum $P_v(k)$ is proportional to the matter power spectrum $P(k)$ at low redshifts,
\begin{eqnarray}
 \label{eq:Pv}
 P_v(k) &=& (H_0 f)^2 \frac{P(k)}{k^2}.
\end{eqnarray}
Using Eq.~\ref{eq:bulk_k} and Eq.~\ref{eq:Pv}, the mean-squared bulk value of ${\bf u}({\bf x_0};R)$ can be shown to be
\begin{eqnarray}
 \label{eq:RMS}
 \sigma_v^2(R) \equiv \; <{\bf u}^2({\bf x_0};R)>\; = {H_0^2\Omega_{\rm m}^{1.1}\over{2\pi^2}}\int   dk\  P(k)W^2(kR),
\end{eqnarray}
where the average is taken over all spatial positions ${\bf x_0}$.

The squared velocity dispersion within a region of size $R$ centered at {$\bf x_0$} can be similarly defined as
\begin{eqnarray}
 \label{eq:disp_x}
 \sigma^2({\bf x_0};R) = \int   d{\bf x} \  |{\bf v(x)}|^2 F(|{\bf x - x_0}|,R) - |{\bf u}({\bf x_0};R)|^2.
\end{eqnarray}
In Fourier space, the ensamble average of Eq.~\ref{eq:disp_x} over ${\bf x_0}$ becomes
\begin{eqnarray}
 \label{eq:disp}
 \sigma^2(R) \equiv \; <\sigma^2({\bf x_0};R)>\; = {H_0^2\Omega_{\rm m}^{1.1}\over{2\pi^2}}\int   dk\  P(k)\left(1-W^2(kR)\right).
\end{eqnarray}
Using Eq.~\ref{eq:RMS} and Eq.~\ref{eq:disp}, the cosmic Mach number can now be defined as
\begin{eqnarray}
 \label{eq:M}
 M(R) \equiv \; <M^2({\bf x_0};R)>^{1/2}\; = \left( \frac{\sigma_v^2(R)} {\sigma^2(R)} \right) ^{1/2}.
\end{eqnarray}
 
As discussed in literature (\citealt{OstSut90}; \citealt*{SutCenOst92, StrCenOst93}), the cosmic Mach number is essentially a measure of the shape of the matter power spectrum: The \textit{rms} bulk flow $\sigma_v(R)$ gets most of its contribution from scales larger than $R$, while the velocity dispersion $\sigma(R)$ is a measure of the magnitude of velocities on scales smaller than $R$ and gets most contribution from small scales (for more detailed analyses of velocity dispersion see also \citet{Watkins97,BahOh96,BahGraCen94}). Furthermore, the statistic $M$ is expected to be independent of the matter power spectrum normalization -- at least on large scales, where the perturbations are still well described by linear theory and affect both $\sigma_v^2(R)$ and $\sigma^2(R)$ equally. $M$ can be a powerful tool to test not only the $\Lambda$CDM scenario, but also a wide range of cosmologies including models with massive neutrinos. Massive neutrinos suppress the matter power spectrum in a scale dependent way, thereby altering the velocity dispersion much more prominently than the bulk flow. Mach number $M$ provides an easy to interpret technique to distinguish between various cosmological models.

In previous work (\citealt*{WatFelHud09}; \citealt*{FelWatHud10}; \citealt*{AgaFelWat12}) we have dealt with peculiar velocity surveys differently. We developed the `minimum variance' formalism (hereafter MV) to make a clean estimate of the the bulk, shear and the octupole moments of the velocity field as a function of scale using the available peculiar velocity data. Higher moments get contribution from progressively smaller scales. In this paper, instead of isolating the higher moments (shear, octupole etc.), we simply estimate the velocity dispersion $\sigma(R)$ that gets contribution from all scales smaller than $R$. In general, the MLE scheme we employ here attempts to minimize the error given a particular survey. The downside of the MLE formalism is the difficulty to directly compare results from different surveys. Since each survey samples the volume differently and although the large scale signal is similar across surveys, the small scale noise is unique. That leads to complicated biases or aliasing that are survey dependent (see \citet{WatFel95,HudColDre00}). The MV formalism corrects for small scale aliasing by using a minimization scheme that treats the volume of the surveys rather than the particular way the survey samples the volume, thus eliminating aliasing and allowing for direct comparison between surveys. 

\section{N-body Simulations}
\label{sec:SIM}

In order to study the statistical distribution of the cosmic Mach number we extract mock surveys from the 41 numerical realizations of a $\Lambda$CDM universe. The N-body simulation we use in our analysis is Large Suite of Dark Matter Simulations (LasDamas, hereafter LD) (\citealt{LasDamas}; McBride et al. 2011 in prep\footnote{http://lss.phy.vanderbilt.edu/lasdamas}). The LD simulation is a suite of 41 independent realizations of dark matter N-body simulations named \textit{Carmen} and have information at redshift $z=0.13$. Using the Ntropy framework \citep{GarConMcB07}, where bound groups of dark matter particles (halos) are identified with a parallel friends-of-friends (FOF) code \citep{FOF}. The cosmological parameters and the design specifications of the LD-\textit{Carmen} are listed in Table~\ref{tab:parameters}.

\begin{table}
\caption{The cosmological parameters and the design specifications of the LD-\textit{Carmen} simulations.}
\begin{tabular}{ll}
 \hline
 \multicolumn{1}{c}{Cosmological parameters}& \multicolumn{1}{c}{LD-\textit{Carmen}} \\ \vspace{-2mm} \\ 
Matter density,						$\Omega_{\rm m}$				& 0.25		\\
Cosmological constant density,		$\Omega_\Lambda$				& 0.75		\\
Baryon density,						$\Omega_{\rm b}$				& 0.04 		\\
Hubble parameter,					$h$ (100 km s$^{-1}$ Mpc$^{-1}$)	& 0.7			\\
Amplitude of matter density fluctuations,	$\sigma_8$					& 0.8			\\
Primordial scalar spectral index,		$n_{\rm s}$					& 1.0			\\ \vspace{-2mm} \\ 
 \multicolumn{1}{c}{Simulation design parameters} & \\ \vspace{-2mm} \\ 
 Simulation box size on a side (\hmpc)								& 1000		\\
 Number of CDM particles											& 1120$^3$	\\
 Initial redshift, $z$												& 49			\\
 Particle mass, $m_{\rm p}$ ($10^{10}$ \hMsun)						& 4.938		\\
 Gravitational force softening length, $f_\epsilon$ (\hkpc)					& 53			\\ \hline
   \end{tabular}
\label{tab:parameters}
\end{table}

We extract 100 mock catalogues from each of the 41 LD-\textit{Carmen} boxes, for a total of 4100 mocks. The mocks are randomly centered inside the boxes. They are extracted to mimic the radial distribution of the real catalogues (described in Sec.~\ref{sec:REAL}), as closely as possible.


\section{Peculiar Velocity Catalogues}
\label{sec:SURVEYS}

\subsection{Real Catalogues}
\label{sec:REAL}

We use a compilation of five galaxy peculiar velocity surveys to study the Mach statistic. This compilation, which we label `DEEP', includes 103 Type-Ia Supernovae (SNIa) \citep{TonSchBar03}, 70 Spiral Galaxy Culsters (SC) Tully-Fisher (TF) clusters \citep{GioHaySal98, DalGioHay99}, 56 Streaming Motions of Abell Clusters (SMAC) fundamental plane (FP) clusters \citep{HudSmiLuc99, HudSmiLuc04}, 50 Early-type Far galaxies (EFAR) FP clusters \citep{ColSagBur01} and 15 TF clusters \citep{Wil99b}. In all, the DEEP catalogue consists of 294 data points. In Fig.~\ref{fig:DEEP}, top row, we show the DEEP catalogue (left-hand panel) and its radial distribution (right-hand panel). The bottom row shows a typical mock extracted from the LD simulations. The procedure to extract mocks is described in Sec.~\ref{sec:MOCKS}.

\subsection{Mock Catalogues}
\label{sec:MOCKS}

Inside the N-body simulation box, we first select a point at random. Next, we extract a mock realization of the real catalogue by imposing the constraint that the mock should have a similar radial distribution to the real catalogue. We do not constrain the mocks to have the same angular distribution as the real catalogue for two reasons: (i) the LD simulation boxes are not dense enough to give us mocks that are exact replicas of the real catalogue, (ii) the objects in a real survey are typically weighted depending only on their velocity errors. Consequently, even though the real catalogue and its mocks have similar radial profiles, their angular distribution differ considerably, with the mocks having a relatively featureless angular distribution. To make the mocks more realistic, we impose a $10^\degree$ latitude zone-of-avoidance cut.

\begin{figure}
     \includegraphics[width= \columnwidth]{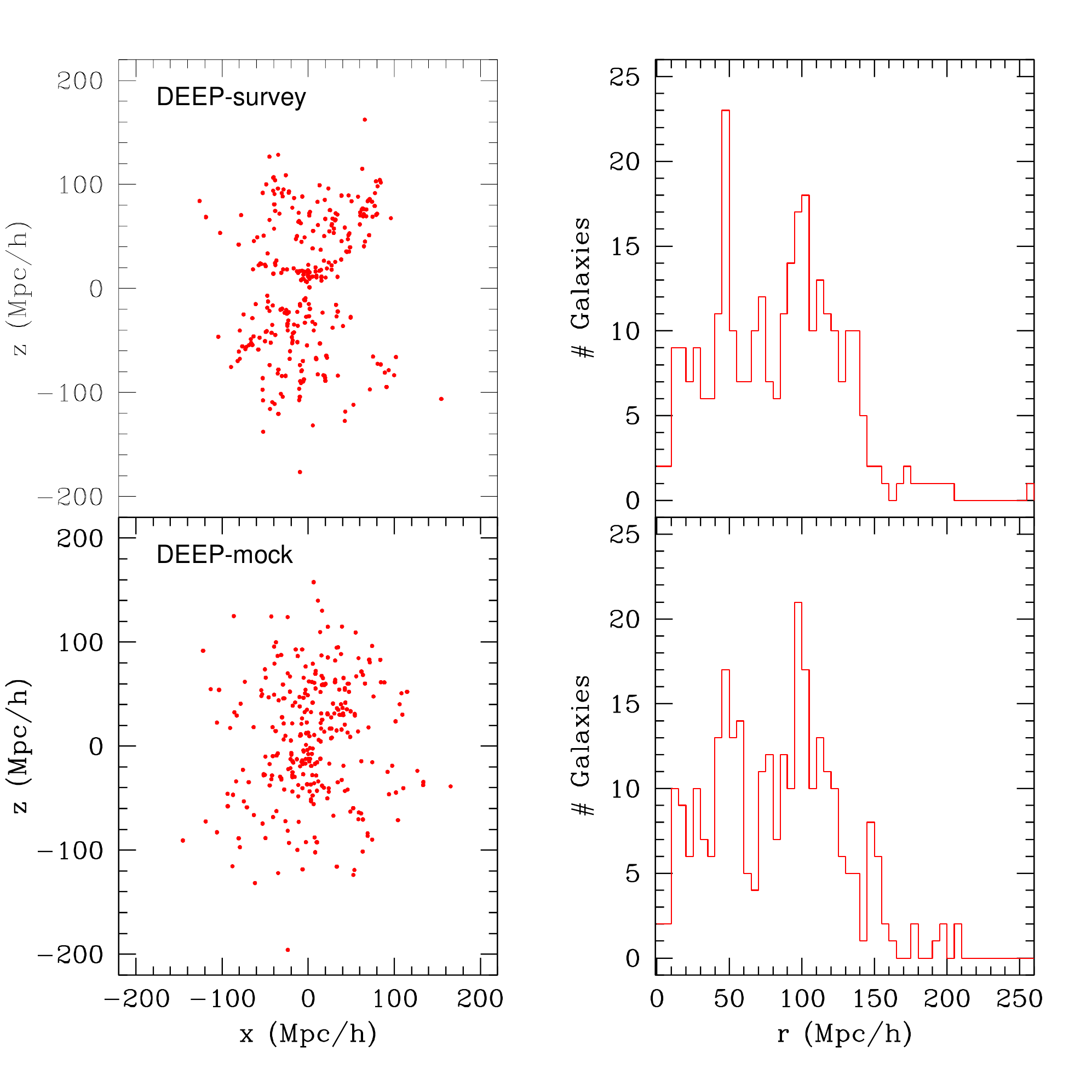}
        \caption{\small{Top row: DEEP catalogue (left) and its radial distribution (right). Bottom row: DEEP mock catalogue (left) and its radial distribution (right).
        }}
    \label{fig:DEEP}
\end{figure}

Using the angular position \{$\hat r_{x}, \hat r_{y}, \hat r_{z}$\}, the true radial distance $d_s$ from the mock center and the peculiar velocity vector ${\bf v}$, we calculate the true line-of-sight peculiar velocity $v_s$ and the redshift $cz=d_s+v_s$ for each mock galaxy (in km/s). We then perturb the true radial distance $d_s$ of the mock galaxy with a velocity error drawn from a Gaussian distribution of width equal to the corresponding real galaxy's velocity error, $e$. Thus, $d_p=d_s+\delta_d$, where $d_p$ is the perturbed radial distance of the mock galaxy (in km/s) and $\delta_d$ is the velocity error drawn from a Gaussian of width $e$. The mock galaxy's measured line-of-sight peculiar velocity $v_p$ is then assigned to be $v_p=cz-d_p$, where $cz$ is the redshift we found above. This procedure ensures that the weights we assign to the mock galaxies are similar to the weights of the real galaxies. In Fig.~\ref{fig:DEEP} we show the angular (left panels) and radial (right panels) distribution of galaxies in the DEEP catalogue (top) and its mock (bottom). 


\section{The Maximum Likelihood Estimate Method}
\label{sec:MLE}

One of the most common weighting scheme used in the analysis of the bulk flow is the maximum likelihood estimate (MLE) method, obtained from a maximum likelihood analysis introduced by \citet{Kai88}. The motion of galaxies is modeled as being due to a streaming flow with Gaussian distributed measurement uncertainties. Given a peculiar velocity survey, the MLE estimate of its bulk flow is obtained from the likelihood function 
\begin{equation}
\label{eq:like}
 L[u_i|\{S_n,\sigma_n,\sigma_*\}]= \prod_n {1\over \sqrt{\sigma_n^2 + \sigma_*^2}}\exp\left( -{1\over 2} {(S_n - \hat r_{n,i}}u_i)^2\over \sigma_n^2 + \sigma_*^2\right),
 \end{equation}
where ${\bf\hat r}_n$ is the unit position vector of the $n$th galaxy; $\sigma_n$ is the measurement uncertainty of the $n$th galaxy; and $\sigma_*$ is the 1-D velocity dispersion accounting for smaller-scale motions. The three components of the bulk flow $u_i$ can be written as weighted sum of the measured radial peculiar velocities of a survey
\begin{equation}
u_i= \sum_n w_{i,n} S_n ,
\label{eq:ui}
\end{equation}
where $S_n$ is the radial peculiar velocity of the $n$th galaxy of a survey; and $w_{i,n}$ is the weight assigned to this velocity in the calculation of $u_i$. Throughout this paper, subscripts $i, j$ and $k$ run over the three spacial components of the bulk flow, while subscripts $m$ and $n$ run over the galaxies. Maximizing the likelihood given by Eq.~\ref{eq:like}, gives the three components of the bulk flow $u_i$ with the MLE weights
\begin{equation}
\label{eq:WEIGHTS}
w_{i,n} = \sum_{j=1}^3 A_{ij}^{-1}{\hat r_{n,j}\over \sigma_n^2 + \sigma_*^2} ,
\end{equation}
where
\begin{equation}
A_{ij}= \sum_n {\hat r_{n,i}\hat r_{n,j}\over \sigma_n^2 + \sigma_*^2} .
\end{equation}

The 1-D velocity dispersion $\sigma_*$ is $1/\sqrt3$ of the 3-D velocity dispersion (see Eq.~\ref{eq:disp}) which we aim to ultimately measure. Since the weights $w_{i,n}$ (and $u_i$) are themselves a function of $\sigma_*$, we converge on to the MLE estimate for $\sigma_*$ iteratively. See \cite*{StrCenOst93} for a discussion on how to estimate the best-fit $u_i$ and $\sigma_*$.

The effective depth of a survey can be roughly estimated by a weighted sum $\sum w_n r_n / \sum w_n$ of the radial distances $r_n$ of the survey objects, where $w_n=1/(\sigma_n^2 + \sigma_*^2)$. This weighting scheme has been used by \cite*{MaOstZha12} in their analyses of peculiar velocity datasets. A drawback of using weights $w_n=1/(\sigma_n^2 + \sigma_*^2)$ in estimating the depth of a survey is that while the weights $w_n$ take into account the measurement errors $\sigma_n$, they do not make any corrections for the survey geometry. A better estimate of the effective depth can be made by looking at the survey window functions $W^2_{ij}$. Window function gives an idea of the scales that contribute to the bulk flow estimates. Ideally, the window function should fall quickly to zero for scales smaller than that being studied. This ensures that the bulk flow estimates are minimally biased from small-scale nonlinearities.

Armed with the MLE weights $w_{i,n}$ from Eq.~\ref{eq:WEIGHTS}, the angle-averaged tensor window function $W^2_{ij}(k)$ (equivalent to $W^2(kR)$ of Eq.~\ref{eq:RMS}) can be constructed (for details, see \citealt*{FelWatHud10}) as
\begin{eqnarray}
 \label{eq:Wij}
 W^2_{ij}(k) &=& \sum_{m,n} w_{i,m} w_{j,n} \int {d^2{\hat k}\over 4\pi} \left( {\bf \hat r}_m\cdot {\bf \hat k} \right)\left( {\bf \hat r}_n\cdot {\bf \hat k} \right) \\ \nonumber
 &&\times\exp \left(ik\ {\bf \hat k}\cdot ({\bf r}_m - {\bf r}_n)\right).
\end{eqnarray}

The diagonal elements $W^2_{ii}$ are the window functions of the bulk flow components $u_i$. The window function gives an idea of the scales that contribute to the bulk flow estimates. Ideally, the window function should fall quickly to zero for scales smaller than that being studied. This ensures that the bulk flow estimates are minimally biased from small-scale nonlinearities. See the MLE \citep{SarFelWat07} and MV \citep{WatFelHud09} window functions of the bulk flow components for a range of surveys.

Having constructed the survey window functions $W^2_{ii}$, the effective depth of the survey can be defined to be the one for which $W^2_{ii}$ is a close match to the window function for an idealized survey.  In order to construct the ideal window functions, we first imagine an idealized survey containing radial velocities that well sample the velocity field in a region.   This survey consists of a large number of objects, all with zero measurement uncertainty.   The radial distribution of this idealized survey is taken to have a Gaussian profile of the form $f(r) \propto e^{-r^2/2R^2}$, where $R$ gives a measure of the depth of the survey.   This idealized survey has easily interpretable bulk flow components that are not affected by small-scale aliasing and which reflect the motion of a well-defined volume.

The MLE weights of an ideal, isotropic survey consisting of $N^\prime$ exact radial velocities $v_{n^\prime}$ measured at randomly selected positions ${\bf r}_{n^\prime}^\prime$ are
\begin{eqnarray}
w^\prime_{i,n^\prime} =  \sum_{j=1}^3 A_{ij}^{-1} {{{\hat r}_{n^\prime,j}^\prime} \over N^\prime},
\end{eqnarray}
where
\begin{eqnarray}
A_{ij}= \sum_{n^\prime=1}^{N^\prime} {{\hat r}_{n^\prime,i}^\prime {\hat r}_{n^\prime,j}^\prime \over N^\prime}.
\end{eqnarray}

Similar to Eq.~\ref{eq:Wij}, the window functions $^{\textit{I}}W^2_{ij}$ for an idealized survey of scale $R$ can be constructed as
\begin{eqnarray}
 \label{eq:Widealij}
 ^{\textit{I}}W^2_{ij}(k;R) &=& \sum_{m,n} w^\prime_{i,m^\prime} w^\prime_{j,n^\prime} \int {d^2{\hat k}\over 4\pi} \left( {\bf \hat r^\prime}_m\cdot {\bf \hat k} \right)\left( {\bf \hat r^\prime}_n\cdot {\bf \hat k} \right) \\ \nonumber
 &&\times\exp \left(ik\ {\bf \hat k}\cdot ({\bf r^\prime}_m - {\bf r^\prime}_n)\right).
\end{eqnarray}

\begin{figure}
     \includegraphics[width= \columnwidth]{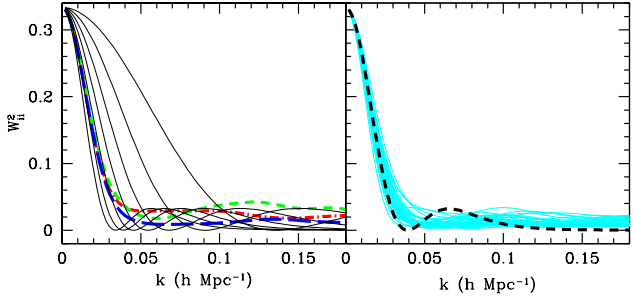}
        \caption{\small{{\bf Left Panel:} The window functions $W^2_{ii}$ of the bulk flow components calculated using MLE weights for the DEEP catalogue. The $x, y, z$ components are dot-dashed, short-dashed and long-dashed lines, respectively. The solid lines are the ideal window functions $^{\textit{I}}W^2_{ij}$ for scales $R=10-40$\hmpc\ (in $5$\hmpc\ increments), the window functions being narrower for larger scales. {\bf Right Panel:} The window functions for a subset of the 4100 DEEP mocks (solid lines). The characteristic depth of the DEEP catalogue and its mocks is $\sim R=35$\hmpc\ (dashed line).
        }}
    \label{fig:WF_DEEP}
\end{figure}

\begin{figure*}
     \includegraphics[width=16.cm]{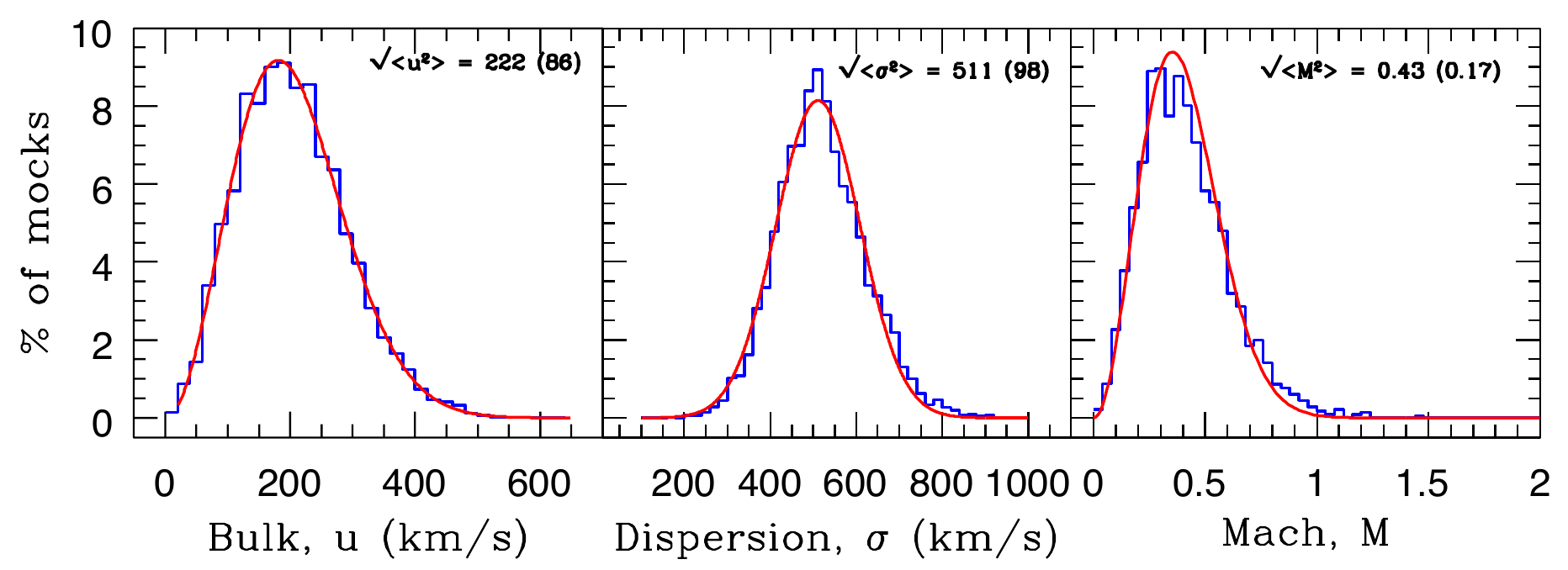}
        \caption{\small{Histograms showing the normalized probability distribution for the 4100 DEEP mocks: bulk flow $u$ (left-hand panel), dispersion $\sigma$ (middle panel) and the cosmic Mach number $M$ (right-hand panel). We also superimpose the best-fitting Maxwellian (for bulk and Mach) and Gaussian (for dispersion) distributions with the same  widths as the corresponding histograms. The \textit{rms} values and the $1\sigma$ CL intervals are mentioned within each panel. These results correspond to the LD cosmology (see Table~\ref{tab:parameters}).
        }}
    \label{fig:v_dist_MLE_DEEP}
\end{figure*}

\begin{figure*}
     \includegraphics[width=16.cm]{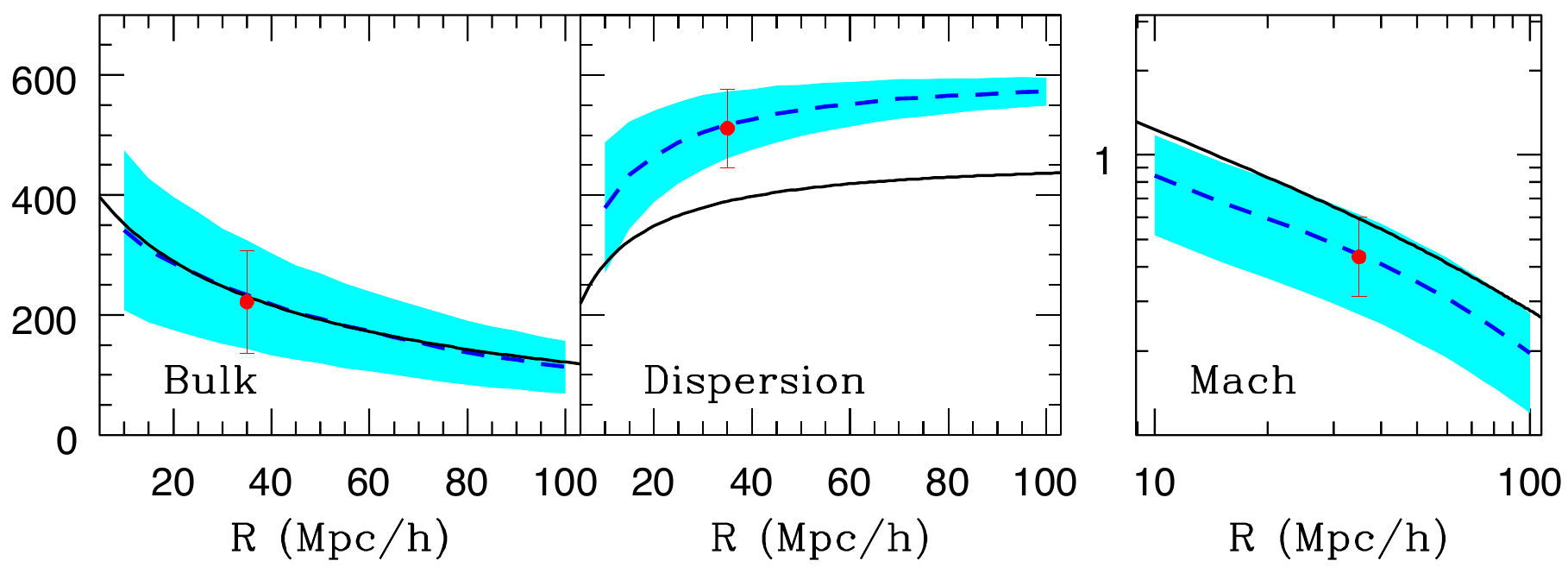}
        \caption{\small{The \textit{rms} values of the bulk flow (left-hand panel), dispersion (middle panel) and the cosmic Mach number (right-hand panel) are plotted as a function of scale $R$. In each panel, the dashed line corresponds to measurements from the Gaussian realizations, whereas the shaded region is the $1\sigma$ CL interval. The solid circle at $R=35$\hmpc\ is the result for the DEEP mocks. The error bar is the statistical variance of the mean calculated from the 4100 DEEP mocks. Linear theory predictions are shown by the solid line. These results correspond to the LD cosmology. The nonlinear contributions to the dispersion are clearly seen in both the center and right panels.
        }}
    \label{fig:MACH_DEEPmocks}
\end{figure*}

In Fig.~\ref{fig:WF_DEEP}, left-hand panel, we show the diagonal window functions $W^2_{ii}$ (see Eq.~\ref{eq:Wij}) of the bulk flow components calculated using MLE weights (see Eq.~\ref{eq:WEIGHTS}) for the DEEP catalogue. The $x, y, z$ components are dot-dashed, short-dashed and long-dashed lines, respectively. Also shown are the ideal window functions $^{\textit{I}}W^2_{ij}$ (see Eq.~\ref{eq:Widealij}) for scales $R=10-40$\hmpc\ (in $5$\hmpc\ increments), the window functions being narrower for larger scales. Comparing the DEEP and the ideal window functions gives the DEEP catalogue an effective depth of $\sim R=35$\hmpc. We note that the weighted sum $\sum w_n r_n / \sum w_n$ gives the DEEP catalogue a depth of 59 \hmpc, an over-estimation by nearly $70\%$. Estimating the survey depth correctly is crucial when it comes to comparing the survey bulk flow with theoretical predictions. One might have a high-quality survey but a poorly estimated depth which can introduce substantial errors when comparing with theory. Throughout this paper, we define the characteristic depth $R$ of a survey as the one estimated from its window functions. The right-hand panel of Fig.~\ref{fig:WF_DEEP} shows the window functions for a subset of the 4100 DEEP mocks (solid lines) extracted from the LD-\textit{Carmen} simulations. The fact that the mock window functions are nearly centered on the $\sim R=35$\hmpc\ ideal window, shows that our procedure for mock extraction works well.

\section{Cosmic Mach Number Statistics}
\label{sec:STATS}

\subsection{Mach statistics for DEEP mocks}
\label{sec:MOCK_STATS}

Using the MLE weighting scheme (Sec.~\ref{sec:MLE}), we estimated the bulk flow moments  \{$u_x, u_y, u_z$\}, the velocity dispersion $\sigma$ and the cosmic Mach number $M$ for each of the 4100 DEEP mock realizations. In Fig.~\ref{fig:v_dist_MLE_DEEP}, we show the probability distribution for the 4100 DEEP mocks: bulk flow $u$ (left-hand panel), dispersion $\sigma$ (middle panel) and the cosmic Mach number $M$ (right-hand panel). We found the \textit{rms} bulk flow to be $\sigma_v=222\pm86$ \kmSec with a  velocity dispersion of $\sigma=511\pm98$ \kmSec. Together this implies $M=0.43\pm0.17$ at $1\sigma$ CL. Since the DEEP mocks have a characteristic depth of $R=35$\hmpc , we can say that for the LD cosmology, the expected Mach number on scales of $R=35$\hmpc\ is $M=0.43\pm0.17$.

\subsection{Mach statistics for Gaussian Realizations}
\label{sec:GAUSSIAN_STATS}

\begin{table}
\centering
\caption{The \textit{rms} values of the bulk flow (2nd column), velocity dispersion (3rd column) and cosmic Mach number (4th column) together with their $1\sigma$ CL intervals for Gaussian windows with width $R$ (1st column). These values are calculated for the LD cosmology (for the LD parameters, see Table~\ref{tab:parameters}).}
\begin{tabular}{|c|c|c|c|}
 \hline
 \multicolumn{1}{c}{$R$} &  \multicolumn{1}{c}{$\sqrt{<u^2>}$} &  \multicolumn{1}{c}{$\sqrt{<\sigma^2>}$} &  \multicolumn{1}{c}{$\sqrt{<M^2>}$} \\
 \multicolumn{1}{c}{(\hmpc)}	& \multicolumn{1}{c}{(\kmSec)}	& \multicolumn{1}{c}{(\kmSec)}	&	\\  
 \hline
10  		& 341  $\pm$ 133	& 379  $\pm$  108 &  0.85  $\pm$  0.33	\\  
15  		& 308  $\pm$ 120	& 433  $\pm$  89 &  0.68  $\pm$  0.27	\\  
20  		& 286  $\pm$ 111	& 464  $\pm$  76 &  0.59  $\pm$  0.23	\\  
25  		& 267  $\pm$ 104	& 487  $\pm$  68 &  0.53  $\pm$  0.21	\\  
30  		& 248  $\pm$ 96	& 504  $\pm$  62 &  0.48  $\pm$  0.19	\\  
35  		& 234  $\pm$ 91	& 517  $\pm$  56 &  0.44  $\pm$  0.17	\\  
40  		& 218  $\pm$ 85	& 526  $\pm$  50 &  0.41  $\pm$  0.16	\\  
45  		& 204  $\pm$ 79	& 535  $\pm$  47 &  0.38  $\pm$  0.15	\\  
50  		& 194  $\pm$ 75	& 541  $\pm$  43 &  0.35  $\pm$  0.14	\\  
55		& 182  $\pm$ 71	& 547  $\pm$  40 &  0.33  $\pm$  0.13	\\  
60  		& 173  $\pm$ 67	& 551  $\pm$  37 &  0.31  $\pm$  0.12	\\  
65  		& 163  $\pm$ 63	& 556  $\pm$  35 &  0.29  $\pm$  0.11	\\  
70  		& 154  $\pm$ 60	& 560  $\pm$  33 &  0.27  $\pm$  0.11	\\  
75		& 145  $\pm$ 57	& 562  $\pm$  31 &  0.26  $\pm$  0.10	\\  
80  		& 137  $\pm$ 53	& 565  $\pm$  29 &  0.24  $\pm$  0.09	\\  
85  		& 130  $\pm$ 51	& 567  $\pm$  27 &  0.23  $\pm$  0.09	\\  
90  		& 125  $\pm$ 48	& 569  $\pm$  26 &  0.22  $\pm$  0.08	\\  
95		& 118  $\pm$ 46	& 571  $\pm$  25 &  0.21  $\pm$  0.08	\\  
100  		& 113  $\pm$ 44	& 572  $\pm$  23 &  0.20  $\pm$  0.07	\\ \hline 
\end{tabular}
\label{tab:SUMMARY}
\end{table}

\begin{figure*}
     \includegraphics[width=16.cm]{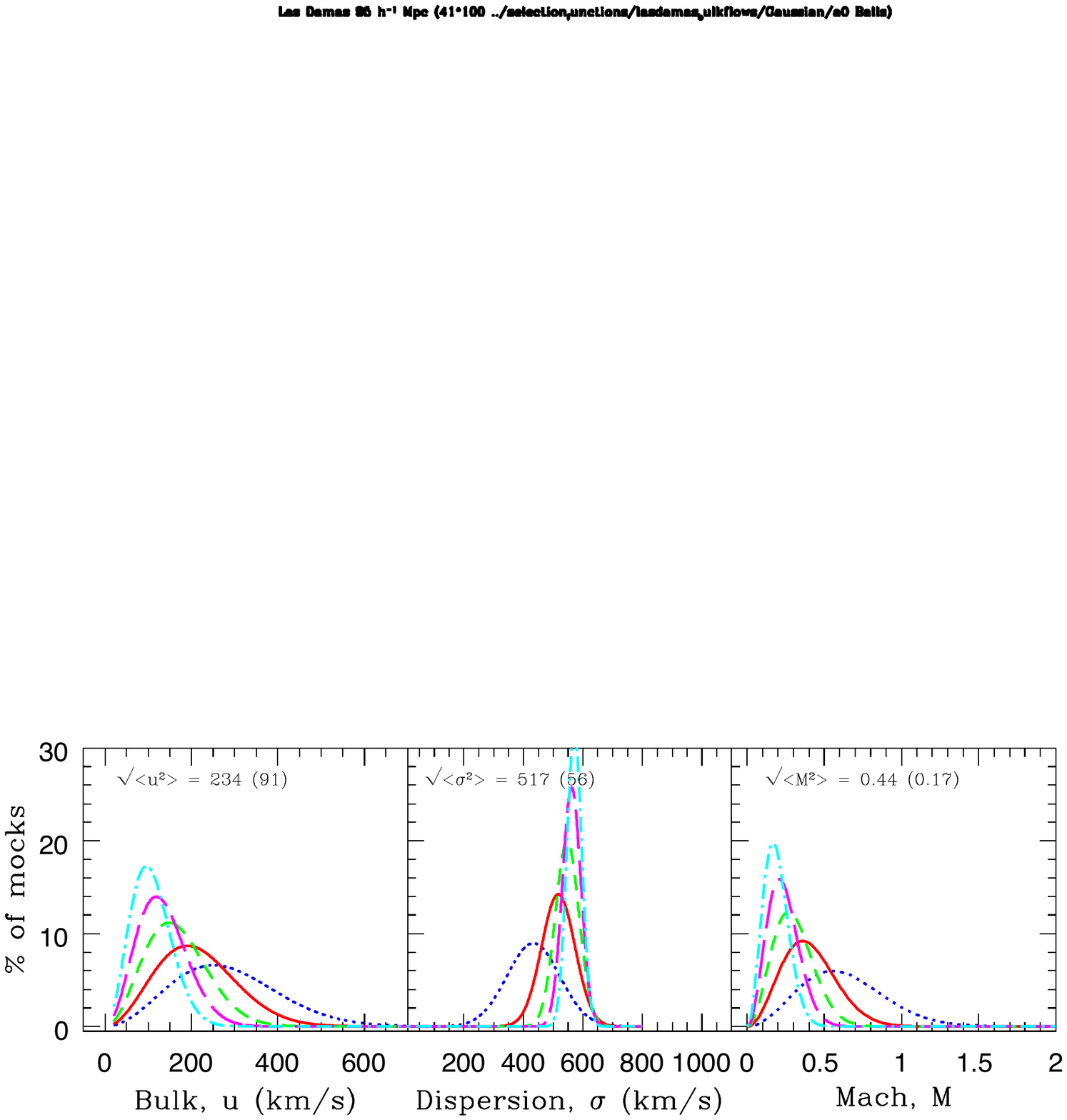}
        \caption{\small{The same as Fig.~\ref{fig:v_dist_MLE_DEEP}, but for Gaussian window with $R=15$\hmpc\ (dotted), $R=35$\hmpc\ (solid), $R=55$\hmpc\ (short-dashed), $R=75$\hmpc\ (long-dashed) and $R=95$\hmpc\ (dot-dashed). For clarity, instead of the histograms, only the best-fitting Maxellian/Gaussian distributions with the same widths as the corresponding histograms are shown. The \textit{rms} values and the $1\sigma$ CL intervals for $R=35$\hmpc\ are listed within each panel, and are in good agreement with the corresponding values for the DEEP mocks (shown in Fig.~\ref{fig:v_dist_MLE_DEEP}). Table~\ref{tab:SUMMARY} summarizes the results for Gaussian widths $R=10-100$ \hmpc.
        }}
    \label{fig:v_dist_gaussian}
\end{figure*}

In order to find the expected Mach number as a function of scale $R$ for the LD cosmology, we went to the same central points for each of the 4100 DEEP mocks and computed the weighted average of the velocities of all the galaxies in the simulation box, the weighting function being $e^{-r^2/2R^2}$. We repeated this for a range of scales between $R=10-100$\hmpc\ in increments of 5\hmpc. We summarize the expected values for the bulk, dispersion and Mach number for scales $R=10-100$\hmpc\ in Table~\ref{tab:SUMMARY}. In Fig.~\ref{fig:MACH_DEEPmocks}, we show the expected values for the bulk, dispersion and Mach number (dashed line) together with their $1\sigma$ CL intervals. The corresponding values for the 4100 DEEP mocks are shown by a solid circle at the characteristic scale $R=35$\hmpc .

The expected bulk ($\sigma_v=234\pm94$ \kmSec), dispersion ($\sigma=517\pm56$ \kmSec) and Mach number ($\sigma=0.44\pm17$) for Gaussian window with $R=35$\hmpc\ are in excellent agreement with the corresponding values for the DEEP mocks. This shows that the DEEP catalogue probes scales up to $\sim R=35$\hmpc , and not $R=59$\hmpc\ as one would have inferred from $\sum w_n r_n / \sum w_n$ using the weights $w_n=1/(\sigma_n^2 + \sigma_*^2)$.

Linear theory predictions for the LD cosmology are shown by the solid lines in Fig.~\ref{fig:MACH_DEEPmocks}. The onset of nonlinear growth in structure formation at low redshifts boosts the velocity dispersion, causing linear theory to over-predict the Mach values.

The probability distributions for $u$, $\sigma$ and $M$ from the Gaussian realizations in the LD simulations are plotted in Fig.~\ref{fig:v_dist_gaussian} for a range of Gaussian widths $R$. For clarity, we only show scales $R=15,\,35,\,55,\,75$ and $95$\hmpc.

As expected, the \textit{rms} bulk flow (dispersion) is a declining (increasing) function of scale $R$ (see Figs.~\ref{fig:MACH_DEEPmocks} and \ref{fig:v_dist_gaussian}). This can be readily understood from the ideal window functions in Fig.~\ref{fig:WF_DEEP}. Larger scales have narrower window functions in Fourier space. Only small scale modes ($k \propto 1/R$) contribute to the \textit{rms} bulk flow integral in Eq.~\ref{eq:RMS}, resulting in smaller bulk flow on larger scales. The dispersion integral (see Eq.~\ref{eq:disp}) gets most of its contribution from higher $k-$values ($k > 1/R$) and gradually increases with narrower windows. Similar histogram trends were found by \cite*{SutCenOst92} from numerical simulations of a CDM universe.

\subsection{Mach statistics for other mocks}
\label{sec:OTHER_MOCKS_STATS}

In the following we extend our analysis to include various different peculiar velocity surveys, specifically to show that our results are not dependent on any radial or angular distributions, nor any distinct morphological types. We compared the Gaussian realizations with  mocks (4100 each) created to emulate the radial selection function of the SBF (Surface Brightness Fluctuations) \citep{TonDreBla01}, ENEAR (Early-type Nearby Galaxies) \citep{daCBerAlo00, BerAlodaC02b, WegBerWil03}, SFI++, SNIa and SC peculiar velocity surveys. Note that the SC and SNIa surveys are also part of our DEEP compilation. The SFI++ (Spiral Field I-band) catalogue \citep{sfi1,sfi2,sfierr09} is the densest and most complete peculiar velocity survey of field spirals to date.  We use data from \citealt{sfierr09}. The sample consists of 2720 TF field galaxies (SFI++f) and 736 groups (SFI++g).

\begin{figure*}
  \begin{flushleft}
   \centering
    \begin{minipage}[c]{1.00\textwidth}
      \centering
      \includegraphics[width=12cm, height=10cm]{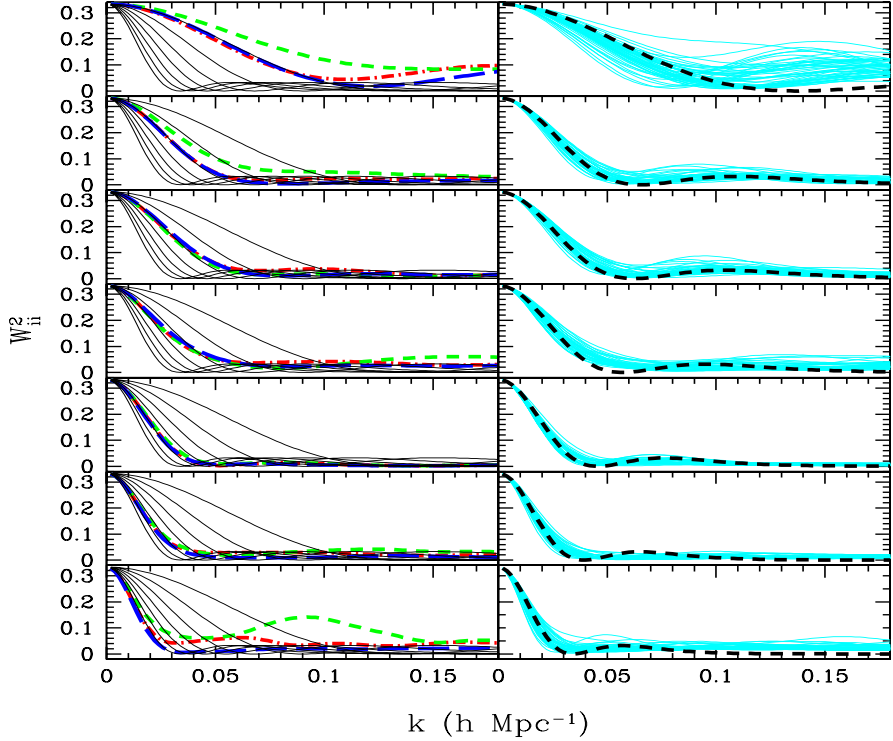}
    \end{minipage}
        \caption{\small{Similar to Fig.~\ref{fig:WF_DEEP}, for the SBF, ENEAR, SFI++g, SNIa, SFI++f, DEEP and SC catalogues (top to bottom row, respectively).
    }}
    \label{fig:WF_ALL}
  \end{flushleft}
\end{figure*}

In Fig.~\ref{fig:WF_ALL}, left-hand panels, we show the window functions $W^2_{ii}$ of the bulk flow components for the SBF, ENEAR, SFI++g, SNIa, SFI++f, DEEP and SC catalogues (top to bottom row, respectively). The right-hand panels show the window functions for a subset of the corresponding mocks. Comparing the window functions of the real catalogues with those of the ideal ones (solid lines in the left-hand panels), we estimate the characteristic depths of the SBF, ENEAR, SFI++g, SNIa, SFI++f, DEEP and SC catalogues to be $R=10,\,19,\,20,\,23,\,30,\,35$ and $40$\hmpc, respectively. The window functions for these depths are shown in the right-hand panels (dashed lines, top to bottom row).

\begin{table*}
\caption{Peculiar velocity statistics for various surveys (1st column). For each survey, 4100 mocks were extracted from the LD cosmology (for the LD parameters, see Table~\ref{tab:parameters}). The characteristic depth $R$ (2nd column) of the mock catalogues is estimated from the effective width of their window functions shown in Fig.~\ref{fig:WF_ALL}. For reference, the error-weighted depth $\sum w_n r_n / \sum w_n$ where $w_n=1/(\sigma_n^2 + \sigma_*^2)$, is listed in the 3rd column. The \textit{rms} values of the bulk flow (4th column), velocity dispersion (5th column) and cosmic Mach number (6th column) together with their $1\sigma$ CL intervals. Columns $7-9$ correspond to the real surveys with the quoted errors calculated using the radial distance uncertainties.}
\begin{tabular}{|l|c|c|c|c|clclclcl}
 \hline
 & & & \multicolumn{3}{c}{Mocks} &  \multicolumn{3}{c}{Real} \\ \vspace{-2mm} \\
 \multicolumn{1}{c}{Survey} &  \multicolumn{1}{c}{$R$} &  \multicolumn{1}{c}{$\frac{\sum w_n r_n}{\sum w_n}$} &  \multicolumn{1}{c}{$\sqrt{<u^2>}$}  &  \multicolumn{1}{c}{$\sqrt{<\sigma^2>}$} &  \multicolumn{1}{c}{$\sqrt{<M^2>}$} &  \multicolumn{1}{c}{$u$}  &  \multicolumn{1}{c}{$\sigma$} &  \multicolumn{1}{c}{$M$} \\
 \vspace{0.1mm} \\
 \multicolumn{1}{c}{} &  \multicolumn{1}{c}{(\hmpc)} &  \multicolumn{1}{c}{(\hmpc)} &  \multicolumn{1}{c}{(\kmSec)}  &  \multicolumn{1}{c}{(\kmSec)} &  \multicolumn{1}{c}{} &  \multicolumn{1}{c}{(\kmSec)}  &  \multicolumn{1}{c}{(\kmSec)} &  \multicolumn{1}{c}{} \\ \vspace{-2mm} \\ \hline

SBF			& 10		& 19		& 322  $\pm$ 125	& 415  $\pm$  100	&  0.74  $\pm$  0.29	& 354  $\pm$  66	& 428  $\pm$  32	&  0.83  $\pm$  0.15		\\ 
ENEAR		& 19		& 34		& 262  $\pm$ 102	& 490  $\pm$  104	&  0.53  $\pm$  0.21	& 292  $\pm$  46	& 528  $\pm$  24	&  0.55  $\pm$  0.09		\\
SFI++g		& 20		& 35		& 280  $\pm$ 101	& 473  $\pm$  66	&  0.59  $\pm$  0.18	& 221  $\pm$  57	& 436  $\pm$  27	&  0.29  $\pm$  0.08		\\  
SNIa			& 23		& 42		& 275  $\pm$ 107	& 465  $\pm$  73	&  0.58  $\pm$  0.21	& 430  $\pm$  87	& 478  $\pm$  47	&  0.90  $\pm$  0.18		\\
SFI++f		& 30		& 52		& 240  $\pm$ 86	& 510  $\pm$  81	&  0.47  $\pm$  0.15	& 320  $\pm$  44	& 503  $\pm$  22	&  0.42  $\pm$  0.06		\\
DEEP  		& 35		& 59		& 222  $\pm$ 86	& 511  $\pm$  65	&  0.43  $\pm$  0.17	& 312  $\pm$  61	& 446  $\pm$  27	&  0.70  $\pm$  0.14		\\
SC  			& 40		& 75		& 227  $\pm$ 88	& 485  $\pm$  43	&  0.47  $\pm$  0.15	& 116  $\pm$  123	& 520  $\pm$  74	&  0.22  $\pm$  0.23		\\ \hline
\end{tabular}
\label{tab:SUMMARY_ALL}
\end{table*}

In Table~\ref{tab:SUMMARY_ALL} we summarize the results for the various surveys (column 1), where the surveys are listed in order of increasing characteristic depth $R$ (column 2) (based on the window functions in Fig.~\ref{fig:WF_ALL}, as described in Sec.~\ref{sec:MLE}). The error-weighted depths $\sum w_n r_n / \sum w_n$, where $w_n=1/(\sigma_n^2 + \sigma_*^2)$ are listed in column $3$, and are typically $\sim\! 75 \%$ larger than $R$.  For the mock surveys, the expected values for the bulk, dispersion and Mach number and their $1\sigma$ CL intervals are summarized in columns $4-6$. Columns $7-9$ are computed using the real surveys. The quoted errors are calculated using the measurement uncertainties $\sigma_n$ of the $n$th galaxy of a survey. Comparing columns $6$ and $9$, the Mach estimates for all catalogues agree at $\sim\!1.5\sigma$ CL for the LD cosmology.

\begin{figure*}
     \includegraphics[width=16.cm]{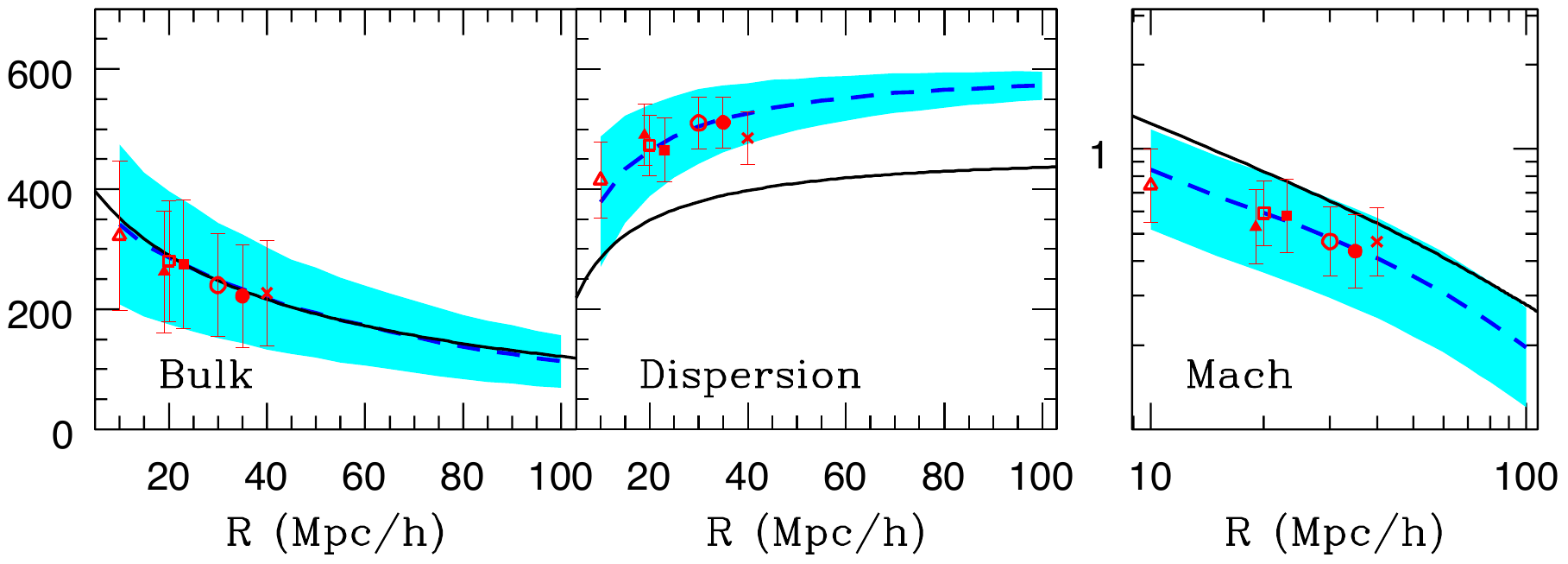}
        \caption{\small{Similar to Fig.~\ref{fig:MACH_DEEPmocks}, including results for the SBF (open triangle), ENEAR (solid triangle), SFI++g (open square), SNIa (solid square), SFI++f (open circle), DEEP (solid circle) and SC (cross) mocks. The DEEP compilation includes the SC, SNIa, SMAC, EFAR and Willick surveys.
        }}
    \label{fig:MACH_MOCKS_L}
\end{figure*}

Similar to Fig.~\ref{fig:MACH_DEEPmocks}, we show results for the SBF, ENEAR, SFI++g, SNIa, SFI++f and SC mocks in Fig.~\ref{fig:MACH_MOCKS_L}. Except for the SBF and SC catalogues, the results for the other catalogues are a close match to their Gaussian counterparts. Our SBF mocks are deeper than the real SBF survey because the LD simulations are not dense enough to extract mocks with depths less than $\twid R=12$\hmpc. This explains why the SBF window functions for the mocks (see Fig.~\ref{fig:WF_ALL}, first row, right-hand panel) are narrower than the one for the SBF's depth of $R=10$\hmpc. Narrower window functions decrease (increase) our bulk flow (dispersion) estimates for the SBF mocks. For the SC mocks, the bulk flow (dispersion) gets excess (suppressed) contribution from smaller scales due to the extended tails of the window functions (see Fig.~\ref{fig:WF_ALL}, row seven). The SC catalogue, with only 70 clusters, does not have a good sky coverage. The DEEP compilation, however, has a much better sky coverage, and the results (see Fig.~\ref{fig:MACH_MOCKS_L}, solid circle) match those from $R=35$\hmpc\ Gaussian mocks. We have included the results for the SBF and SC catalogues to specifically show that if the selection function of the real survey is not properly modeled, the predictions (in our case, based on Gaussian selection function) can be misleading.

For reasonably dense and well sampled velocity surveys, like DEEP, SFI++f, and SFI++g, a close match between the mock and the Gaussian results shows that the Mach analysis for such catalogues is not overly sensitive to the selection functions of the individual mocks. As such, one can skip the step of extracting mock realizations of the observations from N-body simulations, and simply use Mach predictions based on Gaussian selection function $e^{-r^2/2R^2}$ with $R$ set to the characteristic depth of the survey being studied.

\section{Moving Beyond N-Body Simulations: Mach Predictions Using PkANN}
\label{sec:PkANN}

In Sec.~\ref{sec:STATS}, we showed that for velocity surveys with low contamination from small scales, reasonably accurate predictions for the Mach number can be made by extracting mocks having a Gaussian radial profile $e^{-r^2/2R^2}$, $R$ being the characteristic depth of the survey being studied.

A further simplification in the Mach analysis one can hope to achieve is to be able to predict $M(R)$ as a function of scale $R$ without resorting to N-body simulations. Running high-resolution N-body simulations, even in the restricted parameter space around 7-yr $\it{WMAP}$ \citep{Komatsu11} central parameters, is beyond present day computing capabilities. It would be much easier and faster to explore the parameter space using a prescription for the matter power spectrum, and using Eq.~\ref{eq:RMS} and Eq.~\ref{eq:disp} to predict the cosmic Mach number. So far, this has been possible by using linear theory. However, for linear theory results to be applicable, as mentioned in Sec.~\ref{sec:INTRO}, one needs to correct for the nonlinearities in the observed velocity field. Any residual nonlinearity can still bias the Mach predictions.

\begin{figure*}
     \includegraphics[width=16.cm]{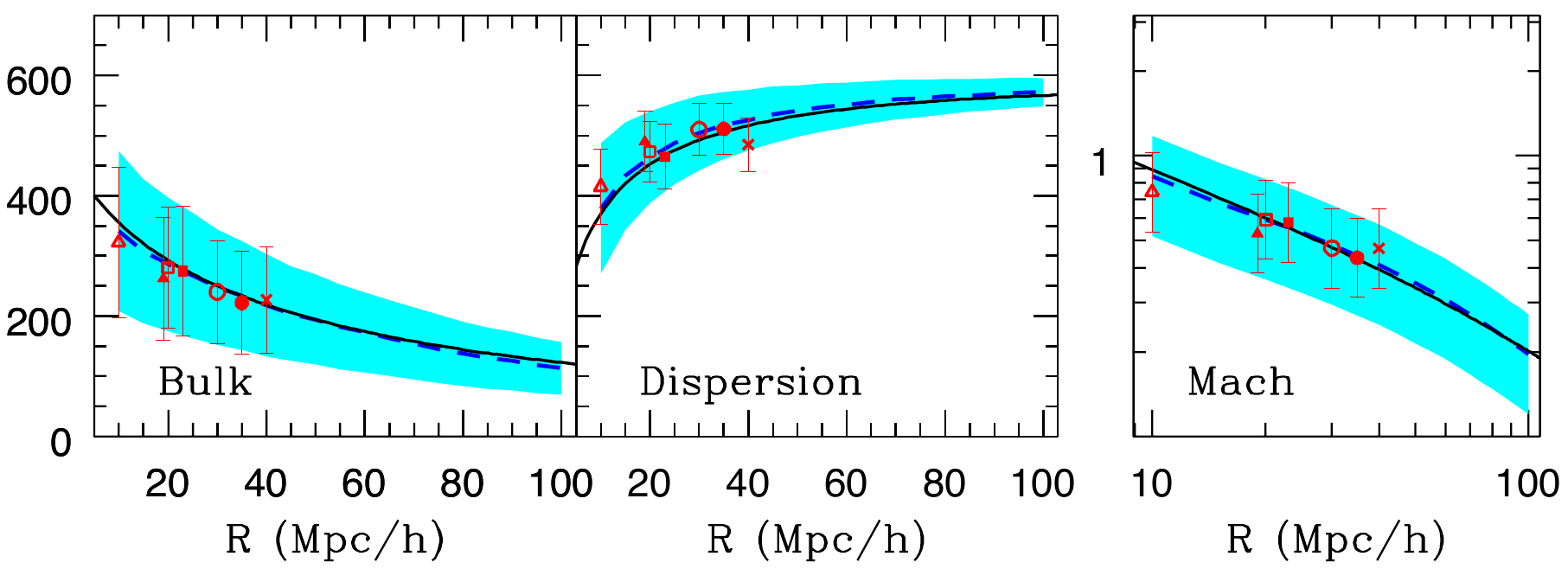}
        \caption{\small{Similar to Fig.~\ref{fig:MACH_MOCKS_L}, but instead of showing linear theory predictions, we plot predictions based on the nonlinear matter power spectrum for the LD cosmology estimated using {\sc PkANN}.
        }}
    \label{fig:MACH_MOCKS_NL}
\end{figure*}

In this section, we attempt to predict $M(R)$ using {\sc PkANN} \citep{AgaAbdFelLahTho12} -- a neural network interpolation scheme to predict the nonlinear matter power spectrum up to $k \ltwid 0.9\,h \textrm{Mpc}^{-1}$ between redshifts $z=0-2$. Although {\sc PkANN} accuracy worsens (starts under-predicting the nonlinear spectrum for $k \gtwid 0.9\,h \textrm{Mpc}^{-1}$), we do not attempt to correct this by smoothing the velocity field over the relevant spatial scale. In Fig.~\ref{fig:MACH_MOCKS_NL}, we replace linear theory predictions shown in Fig.~\ref{fig:MACH_MOCKS_L}, with the ones calculated using {\sc PkANN} for the LD cosmology. {\sc PkANN} (solid lines) gives an excellent match with the N-body results (dashed lines) on all scales, showing {\sc PkANN} can substitute numerical simulations for the purpose of calculating the Mach number given a set of cosmological parameters. Although, we have shown {\sc PkANN}'s performance for only the LD cosmology, it is expected to perform satisfactorily for cosmologies around 7-yr $\it{WMAP}$ central parameters for which {\sc PkANN} has been specifically trained. See \cite{AgaAbdFelLahTho12} for details on the parameter space of {\sc PkANN}'s validity.

\section{Mach Number Estimates From Real Catalogues}
\label{sec:MACH_REAL}

\begin{figure*}
     \includegraphics[width=16.cm]{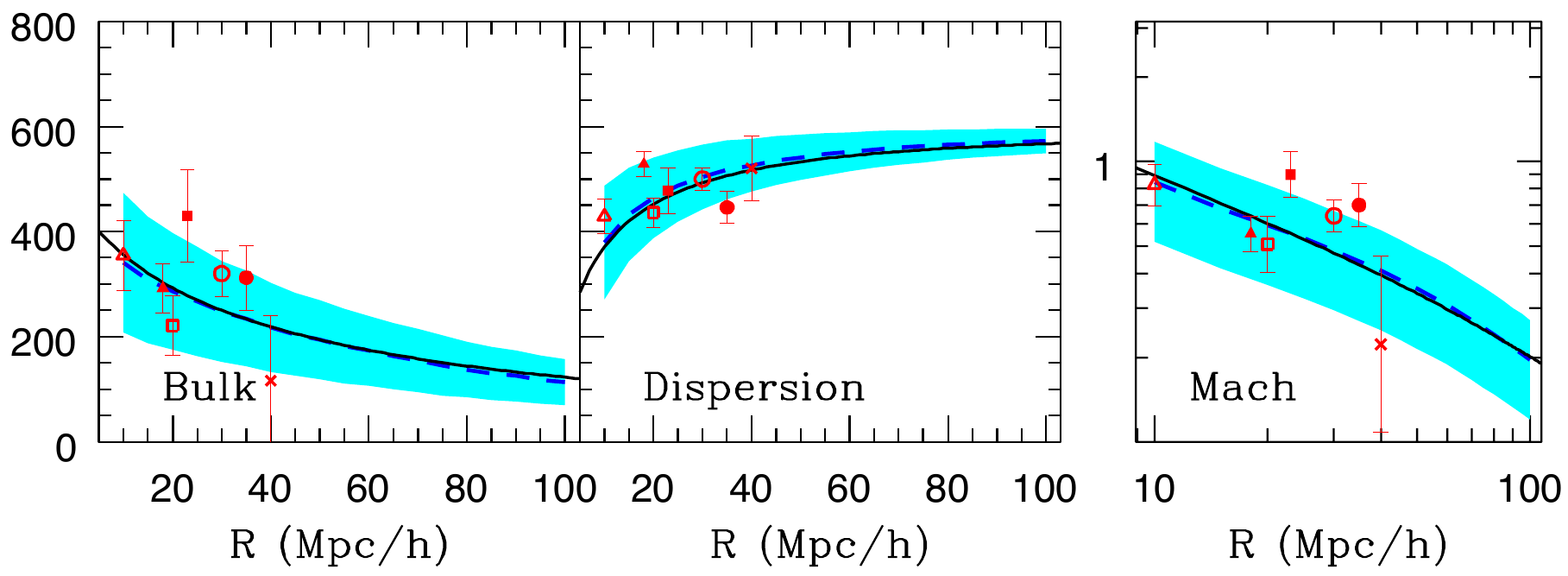}
        \caption{\small{Similar to Fig.~\ref{fig:MACH_MOCKS_NL}, but instead of showing the Mach numbers for the mocks, we plot the Mach numbers for the real surveys. The error bars are calculated using the radial distance uncertainties.
        }}
    \label{fig:MACH_REAL_NL}
\end{figure*}

\cite*{MaOstZha12} measured Mach number for four peculiar velocity surveys (SBF, ENEAR, SNIa and SFI++f ) and found that the $\Lambda$CDM model with 7-yr $\it{WMAP}$ parameters is mildly consistent with the Mach number estimates for these four surveys at $3\sigma$ CL. However, as the authors mention in their work, their estimates are based on using linear approximation for the power spectrum. Given the fact that at low redshifts structure formation has gone nonlinear on small scales, it is necessary to consider nonlinearities when making theoretical predictions. Comparing Figs.~\ref{fig:MACH_MOCKS_L} and \ref{fig:MACH_MOCKS_NL} (middle panels), one can see that dispersion is significantly boosted by nonlinearities, lowering the Mach predictions (third panels) by $1\sigma$ level.

Further, they work with Tophat window functions in their analysis. A Tophat filter assumes a volume-limited survey with a sharp edge in real space. However, the number density of objects sampled in a real survey typically fall at large distances. Real surveys thus have a narrower depth than what a Tophat would suggest. The sharp edge of a Tophat creates both ringing and extended tails in $k$-space. Since it is the small scale modes that are most contaminated by nonlinearities at low redshifts, a Tophat filter leads to aliasing of small-scale power onto larger scales. As such, a Tophat filter is a poor choice if one wants to isolate the contribution from small scales.

It is worth mentioning here that using a Gaussian window function $W^2(kR)=e^{-K^2R^2}$ over-damps the high-$k$ tails associated with a Tophat. The reason being that we only observe the line-of-sight component of the velocity field, whereas the equations presented in Sec.~\ref{sec:MACH} are based on the full 3D velocity measurements. The line-of-sight component extends the tails of the survey window functions in $k$-space (see \citealt{GriPolReyWis87, Kai88}). This is the reason why in our analysis, we do not use $W^2(kR)=e^{-K^2R^2}$; instead, we compute the ideal window functions using only the line-of-sight information (see Eq.~\ref{eq:Widealij}). The extended tails of the ideal window functions can be seen in Fig.~\ref{fig:WF_ALL} and should be contrasted against $W^2(kR)=e^{-K^2R^2}$.

\cite*{MaOstZha12} estimated the characteristic depth of these surveys using $\sum w_n r_n / \sum w_n$ where $w_n=1/(\sigma_n^2 + \sigma_*^2)$. Specifically, they found depths of $16.7,\, 30.5,\, 30.7$ and $50.5$\hmpc\ for the SBF, ENEAR, SNIa and SFI++f, respectively. However, from Fig.~\ref{fig:WF_ALL} and Table~\ref{tab:SUMMARY_ALL} (rows one, two, four and five), we show that these surveys probe scales of $\sim R=10,\,19,\,23$ and $30$\hmpc, respectively. Using linear theory with Tophat filters, and neglecting the survey window functions while estimating the effective depths, makes the bulk flow (and any derived) statistic highly complicated to interpret.

In Sec.~\ref{sec:OTHER_MOCKS_STATS}, we used numerical simulations to study the Mach statistic for SBF, ENEAR, SFI++g, SNIa, SFI++f, DEEP and SC mocks. In this section, we calculated the Mach number using the real catalogues themselves. The results are shown in Fig.~\ref{fig:MACH_REAL_NL} and summarized in Table~\ref{tab:SUMMARY_ALL}, columns $7-9$. We find the Mach observations lie within $\sim\!1.5\sigma$ interval for a $\Lambda$CDM universe with LD parameters. The high uncertainty in the Mach number for the SC catalogue we attribute to its poor sky coverage.

\section{Discussion and conclusions}
\label{sec:Conclude}

The estimates of bulk flow and dispersion on scale $R$ are subject to observational errors stemming from the accuracy levels of distance indicators used, and the survey geometry. Typically, the velocity power spectrum is smoothed using Tophat or Gaussian filters, with results depending on the exact smoothing procedure used. Often, bulk flow results are quoted and inferences drawn about our cosmological model, without paying much attention to the survey window functions which are essential in determining the scales that contribute to quantities derived from peculiar velocities. A statistic such as the cosmic Mach number can be a useful tool to test theories of structure formation, provided the observational uncertainties are accounted for and the scale is properly determined.

In this paper, we studied the statistical distribution of Mach number by extracting mock realizations of the real peculiar velocity catalogues from LasDamas numerical simulations. We showed that the Mach number estimates from the real catalogues agree with the expectations for a $\Lambda$CDM universe at $\sim\!1.5\sigma$ level at the characteristic scales of the surveys. We checked if our Mach expectations derived from mock surveys were biased by the selection function effects:  we extracted realizations with a Gaussian profile $f(r) \propto e^{-r^2/2R^2}$ and found no significant change to our Mach values for the mock surveys.

We compared results from numerical simulations to show that theoretical prediction of Mach number based on linear theory of structure formation is inaccurate. Specifically, small-scale nonlineraties increase velocity dispersion, thereby lowering the Mach predictions by about $1\sigma$ for a $\Lambda$CDM universe with $\it{WMAP}$ type cosmology. We presented an alternative method to study the cosmic Mach number -- by using a prescription for the nonlinear matter power spectrum, instead of running time-consuming and computationally-intensive numerical simulations. Nonlinear power spectrum interpolators like {\sc PkANN} offer tremendous leverage over numerical simulations, by being able to explore the parameter space quickly. The role of such interpolating schemes in the study of quantities derived from peculiar velocities needs further investigation. Also, in the future we plan to employ MV--like formalism to study this statistic, reduce the nonlinear signal to below the statistical errors and thus create a truly linear Mach number statistic that can be used to directly compare results from disparate surveys as a function of the volume chosen and probed.

\section{Acknowledgements}
We are grateful to R\'oman Scoccimarro and the LasDamas collaboration for providing us with the simulations and to Rick Watkins for his thoughtful comments.
This work was supported in part by the National Science Foundation through TeraGrid resources provided by the NCSA.

\bibliographystyle{mn2e}
\bibliography{mach}

\begin{thebibliography}{36}
\expandafter\ifx\csname natexlab\endcsname\relax\def\natexlab#1{#1}\fi

\bibitem[{{Agarwal} {et~al.}(2012{\natexlab{a}}){Agarwal}, {Abdalla},
  {Feldman}, {Lahav}, \& {Thomas}}]{AgaAbdFelLahTho12}
{Agarwal} S., {Abdalla} F.~B., {Feldman} H.~A., {Lahav} O., {Thomas} S.~A.,
  2012{\natexlab{a}}, \mnras, 424, 1409

\bibitem[{{Agarwal} {et~al.}(2012{\natexlab{b}}){Agarwal}, {Feldman}, \&
  {Watkins}}]{AgaFelWat12}
{Agarwal} S., {Feldman} H.~A., {Watkins} R., 2012{\natexlab{b}}, \mnras, 424,
  2667

\bibitem[{{Bahcall} {et~al.}(1994){Bahcall}, {Gramann}, \& {Cen}}]{BahGraCen94}
{Bahcall} N.~A., {Gramann} M., {Cen} R., 1994, \apj, 436, 23

\bibitem[{{Bahcall} \& {Oh}(1996)}]{BahOh96}
{Bahcall} N.~A., {Oh} S.~P., 1996, \apjl, 462, L49

\bibitem[{{Bernardi} {et~al.}(2002){Bernardi}, {Alonso}, {da Costa}, {Willmer},
  {Wegner}, {Pellegrini}, {Rit{\'e}}, \& {Maia}}]{BerAlodaC02b}
{Bernardi} M., {Alonso} M.~V., {da Costa} L.~N., {Willmer} C.~N.~A., {Wegner}
  G., {Pellegrini} P.~S., {Rit{\'e}} C., {Maia} M.~A.~G., 2002, \aj, 123, 2990

\bibitem[{{Colless} {et~al.}(2001){Colless}, {Saglia}, {Burstein}, {Davies},
  {McMahan}, \& {Wegner}}]{ColSagBur01}
{Colless} M., {Saglia} R.~P., {Burstein} D., {Davies} R.~L., {McMahan} R.~K.,
  {Wegner} G., 2001, \mnras, 321, 277

\bibitem[{{da Costa} {et~al.}(2000){da Costa}, {Bernardi}, {Alonso}, {Wegner},
  {Willmer}, {Pellegrini}, {Rit{\'e}}, \& {Maia}}]{daCBerAlo00}
{da Costa} L.~N., {Bernardi} M., {Alonso} M.~V., {Wegner} G., {Willmer}
  C.~N.~A., {Pellegrini} P.~S., {Rit{\'e}} C., {Maia} M.~A.~G., 2000, \aj, 120,
  95

\bibitem[{{Dale} {et~al.}(1999a){Dale}, {Giovanelli}, {Haynes}, {Campusano}, \&
  {Hardy}}]{DalGioHay99}
{Dale} D.~A., {Giovanelli} R., {Haynes} M.~P., {Campusano} L.~E., {Hardy} E.,
  1999a, \aj, 118, 1489

\bibitem[{{Davis} {et~al.}(1985){Davis}, {Efstathiou}, {Frenk}, \&
  {White}}]{FOF}
{Davis} M., {Efstathiou} G., {Frenk} C.~S., {White} S.~D.~M., 1985, \apj, 292,
  371

\bibitem[{{Feldman} \& {Watkins}(1994)}]{FelWat94}
{Feldman} H.~A., {Watkins} R., 1994, \apjl, 430, L17

\bibitem[{{Feldman} {et~al.}(2010){Feldman}, {Watkins}, \&
  {Hudson}}]{FelWatHud10}
{Feldman} H.~A., {Watkins} R., {Hudson} M.~J., 2010, \mnras, 407, 2328

\bibitem[{{Gardner} {et~al.}(2007){Gardner}, {Connolly}, \&
  {McBride}}]{GarConMcB07}
{Gardner} J.~P., {Connolly} A., {McBride} C., 2007, in Astronomical Society of
  the Pacific Conference Series, Vol. 376, Astronomical Data Analysis Software
  and Systems XVI, {Shaw} R.~A., {Hill} F., {Bell} D.~J., eds., p.~69

\bibitem[{{Giovanelli} {et~al.}(1998){Giovanelli}, {Haynes}, {Salzer},
  {Wegner}, {da Costa}, \& {Freudling}}]{GioHaySal98}
{Giovanelli} R., {Haynes} M.~P., {Salzer} J.~J., {Wegner} G., {da Costa} L.~N.,
  {Freudling} W., 1998, \aj, 116, 2632

\bibitem[{{Grinstein} {et~al.}(1987){Grinstein}, {Politzer}, {Rey}, \&
  {Wise}}]{GriPolReyWis87}
{Grinstein} B., {Politzer} H.~D., {Rey} S.-J., {Wise} M.~B., 1987, \apj, 314,
  431

\bibitem[{{Hudson} {et~al.}(2000){Hudson}, {Colless}, {Dressler}, \&
  {Giovanelli}}]{HudColDre00}
{Hudson} M.~J., {Colless} M., {Dressler} A., {Giovanelli} R., 2000, in
  Astronomical Society of the Pacific Conference Series, Vol. 201, Cosmic Flows
  Workshop, {Courteau} S., {Willick} J., eds., p. 159

\bibitem[{{Hudson} {et~al.}(2004){Hudson}, {Smith}, {Lucey}, \&
  {Branchini}}]{HudSmiLuc04}
{Hudson} M.~J., {Smith} R.~J., {Lucey} J.~R., {Branchini} E., 2004, \mnras,
  352, 61

\bibitem[{{Hudson} {et~al.}(1999){Hudson}, {Smith}, {Lucey}, {Schlegel}, \&
  {Davies}}]{HudSmiLuc99}
{Hudson} M.~J., {Smith} R.~J., {Lucey} J.~R., {Schlegel} D.~J., {Davies} R.~L.,
  1999, \apjl, 512, L79

\bibitem[{{Kaiser}(1988)}]{Kai88}
{Kaiser} N., 1988, \mnras, 231, 149

\bibitem[{{Komatsu} {et~al.}(2011){Komatsu}, {Smith}, {Dunkley}, {Bennett},
  {Gold}, {Hinshaw}, {Jarosik}, {Larson}, {Nolta}, {Page}, {Spergel},
  {Halpern}, {Hill}, {Kogut}, {Limon}, {Meyer}, {Odegard}, {Tucker}, {Weiland},
  {Wollack}, \& {Wright}}]{Komatsu11}
{Komatsu} E., {Smith} K.~M., {Dunkley} J., {Bennett} C.~L., {Gold} B.,
  {Hinshaw} G., {Jarosik} N., {Larson} D., {Nolta} M.~R., {Page} L., {Spergel}
  D.~N., {Halpern} M., {Hill} R.~S., {Kogut} A., {Limon} M., {Meyer} S.~S.,
  {Odegard} N., {Tucker} G.~S., {Weiland} J.~L., {Wollack} E., {Wright} E.~L.,
  2011, \apjs, 192, 18

\bibitem[{{Linder}(2005)}]{Lin05}
{Linder} E.~V., 2005, \prd, 72, 043529

\bibitem[{{Ma} {et~al.}(2012){Ma}, {Ostriker}, \& {Zhao}}]{MaOstZha12}
{Ma} Y.-Z., {Ostriker} J.~P., {Zhao} G.-B., 2012, \jcap, 6, 26

\bibitem[{{Masters} {et~al.}(2006){Masters}, {Springob}, {Haynes}, \&
  {Giovanelli}}]{sfi1}
{Masters} K.~L., {Springob} C.~M., {Haynes} M.~P., {Giovanelli} R., 2006, \apj,
  653, 861

\bibitem[{{McBride} {et~al.}(2009){McBride}, {Berlind}, {Scoccimarro},
  {Wechsler}, {Busha}, {Gardner}, \& {van den Bosch}}]{LasDamas}
{McBride} C., {Berlind} A., {Scoccimarro} R., {Wechsler} R., {Busha} M.,
  {Gardner} J., {van den Bosch} F., 2009, in BAAS, Vol.~41, p. 425.06

\bibitem[{{Ostriker} \& {Suto}(1990)}]{OstSut90}
{Ostriker} J.~P., {Suto} Y., 1990, \apj, 348, 378

\bibitem[{{Sarkar} {et~al.}(2007){Sarkar}, {Feldman}, \&
  {Watkins}}]{SarFelWat07}
{Sarkar} D., {Feldman} H.~A., {Watkins} R., 2007, \mnras, 375, 691

\bibitem[{{Springob} {et~al.}(2007){Springob}, {Masters}, {Haynes},
  {Giovanelli}, \& {Marinoni}}]{sfi2}
{Springob} C.~M., {Masters} K.~L., {Haynes} M.~P., {Giovanelli} R., {Marinoni}
  C., 2007, \apjs, 172, 599

\bibitem[{{Springob} {et~al.}(2009){Springob}, {Masters}, {Haynes},
  {Giovanelli}, \& {Marinoni}}]{sfierr09}
---, 2009, \apjs, 182, 474

\bibitem[{{Strauss} {et~al.}(1993){Strauss}, {Cen}, \&
  {Ostriker}}]{StrCenOst93}
{Strauss} M.~A., {Cen} R., {Ostriker} J.~P., 1993, \apj, 408, 389

\bibitem[{{Suto} {et~al.}(1992){Suto}, {Cen}, \& {Ostriker}}]{SutCenOst92}
{Suto} Y., {Cen} R., {Ostriker} J.~P., 1992, \apj, 395, 1

\bibitem[{{Tonry} {et~al.}(2001){Tonry}, {Dressler}, {Blakeslee}, {Ajhar},
  {Fletcher}, {Luppino}, {Metzger}, \& {Moore}}]{TonDreBla01}
{Tonry} J.~L., {Dressler} A., {Blakeslee} J.~P., {Ajhar} E.~A., {Fletcher}
  A.~B., {Luppino} G.~A., {Metzger} M.~R., {Moore} C.~B., 2001, \apj, 546, 681

\bibitem[{{Tonry} {et~al.}(2003){Tonry}, {Schmidt}, {Barris}, {Candia},
  {Challis}, {Clocchiatti}, {Coil}, {Filippenko}, {Garnavich}, {Hogan},
  {Holland}, {Jha}, {Kirshner}, {Krisciunas}, {Leibundgut}, {Li}, {Matheson},
  {Phillips}, {Riess}, {Schommer}, {Smith}, {Sollerman}, {Spyromilio},
  {Stubbs}, \& {Suntzeff}}]{TonSchBar03}
{Tonry} J.~L., {Schmidt} B.~P., {Barris} B., {Candia} P., {Challis} P.,
  {Clocchiatti} A., {Coil} A.~L., {Filippenko} A.~V., {Garnavich} P., {Hogan}
  C., {Holland} S.~T., {Jha} S., {Kirshner} R.~P., {Krisciunas} K.,
  {Leibundgut} B., {Li} W., {Matheson} T., {Phillips} M.~M., {Riess} A.~G.,
  {Schommer} R., {Smith} R.~C., {Sollerman} J., {Spyromilio} J., {Stubbs}
  C.~W., {Suntzeff} N.~B., 2003, \apj, 594, 1

\bibitem[{{Watkins}(1997)}]{Watkins97}
{Watkins} R., 1997, \mnras, 292, L59

\bibitem[{{Watkins} \& {Feldman}(1995)}]{WatFel95}
{Watkins} R., {Feldman} H.~A., 1995, \apjl, 453, L73+

\bibitem[{{Watkins} {et~al.}(2009){Watkins}, {Feldman}, \&
  {Hudson}}]{WatFelHud09}
{Watkins} R., {Feldman} H.~A., {Hudson} M.~J., 2009, \mnras, 392, 743

\bibitem[{{Wegner} {et~al.}(2003){Wegner}, {Bernardi}, {Willmer}, {da Costa},
  {Alonso}, {Pellegrini}, {Maia}, {Chaves}, \& {Rit{\'e}}}]{WegBerWil03}
{Wegner} G., {Bernardi} M., {Willmer} C.~N.~A., {da Costa} L.~N., {Alonso}
  M.~V., {Pellegrini} P.~S., {Maia} M.~A.~G., {Chaves} O.~L., {Rit{\'e}} C.,
  2003, \aj, 126, 2268

\bibitem[{{Willick}(1999)}]{Wil99b}
{Willick} J.~A., 1999, \apj, 522, 647

\end{thebibliography}

\end{document}